\documentclass[%
 reprint,
 amsmath,amssymb,
 aps,
 prl,
]{revtex4-1}

\usepackage{graphicx}
\usepackage{dcolumn}
\usepackage{bm}
\usepackage{float}
\usepackage{color}
\usepackage{soul}
\usepackage{subfigure}
\usepackage{amsmath} 
\newcommand{\angstrom}{\text{\normalfont\AA}}

\begin{document}

\preprint{APS/123-QED}

\title{Critical Strain for Surface Nucleation of Dislocations in Silicon
}

\author{Xiaohan Zhang}
\author{Wei Cai}%
 \email{caiwei@stanford.edu}
\affiliation{%
 Department of Mechanical Engineering, Stanford University\\
}%

\date{\today}

\begin{abstract}
A long-standing discrepancy exists between experiments and atomistic models concerning the critical strain needed for surface nucleation of dislocations in silicon-germanium systems.  While dislocation nucleation is readily observed in hetero-epitaxial thin films with misfit strains less than 4\%, existing atomistic models predict that a critical strain over 7.8\% is needed to overcome the kinetic barrier for dislocation nucleation.
Using zero-temperature energy barrier calculations and finite-temperature Molecular Dynamics simulations, we show that 3-dimensional surface features such as a sharply bent step can lower the predicted critical nucleation strain of a shuffle-set dislocation to 6.4\%, and that of a shuffle-glide dislocation complex to 5.3\%.
Consistent findings are obtained using both the Stillinger-Weber (SW) and modified embedded-atom method (MEAM) potentials, providing support to the physical relevance of the shuffle-glide dislocation complex, which was previously considered as an artifact of the SW potential.
%
%
\end{abstract}

\maketitle


%

%
The introduction of SiGe-Si heteroepitaxial systems to modern transistors is a widely adopted way to boost the electronic performance. The misfit strain generates stress in the transistor channel and allows higher carrier mobility and reduced source/drain resistance~\cite{kim2014strain}.
%
%
However, dislocations tend to nucleate under the high stress, relaxing the beneficial stress and causing electrical shorting~\cite{galle2016dislocation,cea2014high}. 
Predicting the critical strain conditions for dislocation nucleation in Si-based semiconductors is thus important to the fabrication of modern integrated circuits.

%
%
%
%


A long-standing discrepancy exists between experiments and atomistic models concerning the critical strain for surface nucleation of dislocations in Si-Ge systems. 
%
For example, the misfit strain of $4\%$ between Si and Ge is sufficient for dislocation nucleation in a core-shell nanowire~\cite{dayeh2012direct}.
%
Dislocations are also readily observed to nucleate in a SiGe film on a Si substrate (with misfit strain $<4\%$) during annealing~\cite{eaglesham1989dislocation,gillard1993study,noble1989thermal,perovic1995introduction}.
%
However, both molecular dynamics (MD) simulations and energy barrier calculations have predicted very high energy barriers for dislocation nucleation under this level of strain~\cite{maras2016global}.  In order to lower the barriers so that they can be overcome by thermal fluctuations, these atomistic models would predict a critical strain for dislocation nucleation that greatly exceeds the experimental values~\cite{izumi2008dislocation,godet2009dislocation, maras2017atomic}.
%



%

\begin{figure}[H]
\centering
\includegraphics[width=.7\linewidth]{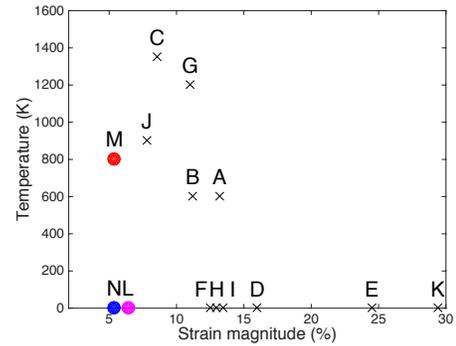}
\caption{Predicted critical strain of dislocation nucleation in Si by atomistic models.
(A-K): results in the literature using the Stillinger-Weber (SW) potential~\cite{godet2009evidence, izumi2008dislocation,shima2010reaction,thaulow2011atomistic,li2010dislocation,godet2004computer,saeed2009transition} and {\it ab initio}~\cite{godet2006dislocation} models.
%
(L,M,N): this work using the SW and MEAM potentials.
See text and {supplementary information} for the explanation of each symbol.
}
\label{fig:literature_summary}
\end{figure}

The discrepancy between experiments and atomistic models of dislocation nucleation may be attributed to several factors, including the artifacts of interatomic potentials, and uncertainties concerning the nucleation mechanisms~\cite{godet2009evidence}, surface geometry and roughness~\cite{huang2008mesoscopic}, and the effect of Ge atom distribution~\cite{bolkhovityanov2012ge}. 
%
%
Here we assume that the presence of Ge atoms is not the main cause of this discrepancy.  This view is supported by the same kind of discrepancy on dislocation nucleation in pure Si.
Fig.~\ref{fig:literature_summary} shows that the critical (applied normal) strain for dislocation nucleation predicted by existing atomistic simulations falls in the range from 7.8\% to 30\%.
However, shuffle-set dislocations have been observed to nucleate in Si under $\sim3\%$ compressive strain~\cite{rabier2010dislocations}, which is comparable to the misfit strain in SiGe/Si epitaxial systems discussed above.
%
%
%
%
%
%
%
%

\begin{figure*}[tp]
\includegraphics[width=.8\linewidth]{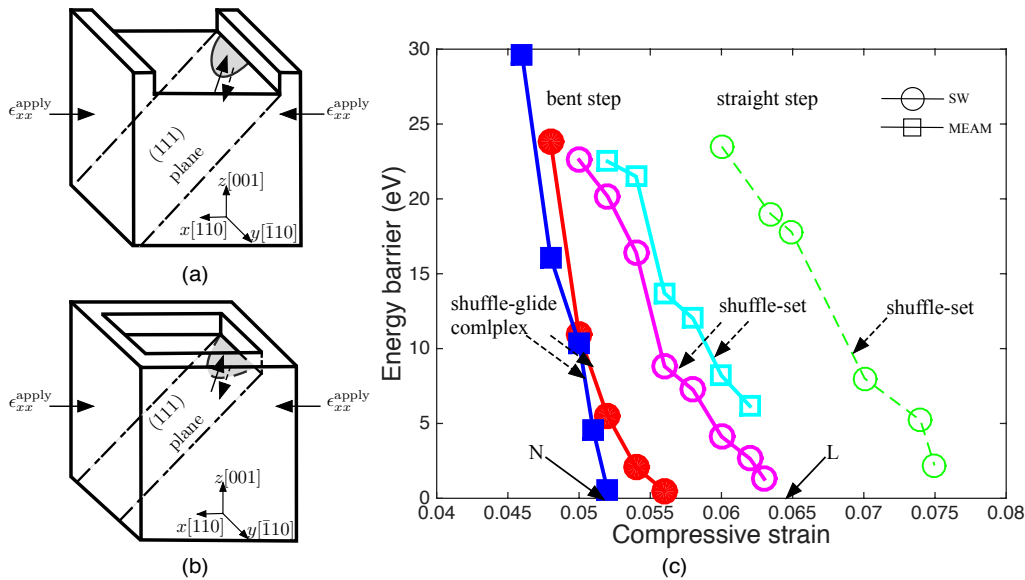}
\caption{Simulation cell for energy barrier calculations for dislocation nucleation from Si surface with (a) straight and (b) sharply bent step structures. $\epsilon^{\text{apply}}$ represents applied compressive strain.  The shaded areas correspond to slipped regions enclosed by the dislocation loop, with the arrows indicating the direction of slip.  (c) Predicted nucleation energy barriers using SW (circles) and MEAM (squares) potentials.  Dashed line: nucleation from straight step; solid line: nucleation from bent step.  Open symbols: shuffle-set dislocation; closed symbols: shuffle-glide dislocation complex.
%
}
\label{fig:neb-data}
\end{figure*}

In this work, we show that the discrepancy in Si can be significantly reduced by considering 3-dimensional surface features.
In particular, our energy barrier calculations predict the nucleation of 
a shuffle-set dislocation from a sharply bent surface step at a critical strain of 6.4\% (point L in Fig.~\ref{fig:literature_summary}).
%
%
Our MD simulations further reveal that a shuffle-glide dislocation complex can nucleate from the bent step at an even lower critical strain of $5.3\%$ (point M), a finding that is subsequently confirmed by energy barrier calculations (point N).
The shuffle-glide dislocation complex was reported earlier (as a micro-twin) which nucleated from a straight surface step at a critical strain of $7.8\%$ (point J), but was suspected to be an artifact of the Stillinger-Weber (SW) potential~\cite{godet2004theoretical}.
Here we show that the same nucleation mechanism occurs in both the SW and the modified embedded-atom method  (MEAM)~\cite{baskes1992modified} potentials of Si, supporting its physical relevance.
The combination of the shuffle-glide dislocation complex and the bent surface step leads to the lowest critical nucleation strain ($5.3\%$) in Si predicted by atomistic models so far, much closer to the experimental value ($\sim3\%$) than before ($7.8\%$).
Fig.~\ref{fig:neb-data}(a) and (b) illustrate the geometry of our atomistic simulation cells containing straight and bent surface steps, respectively.
The free surface is in the $(001)$ orientation (normal to the $z$ axis), and the steps are along the $\langle110\rangle$ directions (along $x$ and $y$).
The step height is $2\,a$, where $a$ is the lattice constant of Si. 
%
The simulation cells are subjected to periodic boundary conditions (PBC) in $x$ and $y$ and free surface boundary conditions in $z$.
Uniaxial compressive strain is applied along $x$ with zero strain along $y$.
The simulation cells for energy barrier calculations have the dimension of $81.4\,\angstrom\,(x)$, $81.4\,\angstrom\,(y)$ and $114\,\angstrom\,(z)$, containing about $14,000$ atoms, while the MD simulation cells have the dimension of $192\,\angstrom\,(x)$, $192\,\angstrom\,(y)$ and $214\,\angstrom\,(z)$, containing about $732,000$ atoms. 
%
To address the possible artifact of interatomic potentials, we use both SW and MEAM potentials for our energy barrier calculations and MD simulations. SW potential has been used extensively in the literature~\cite{godet2009evidence,pizzagalli2013new}, while the MEAM potential has been shown to be more accurate for dislocation related properties, such as the brittle-to-ductile transition in Si~\cite{kang2007brittle,kang2010size,ryu2008comparison}. 
%
We note that proper preparation of the initial and final states to account for surface reconstructions is essential for obtaining consistent results (see {supplementary information}).
%
%


\begin{figure*}[t!]
\begin{minipage}{\linewidth}
\includegraphics[trim={0cm 0cm 0cm 0cm}, clip, width=.8\linewidth]{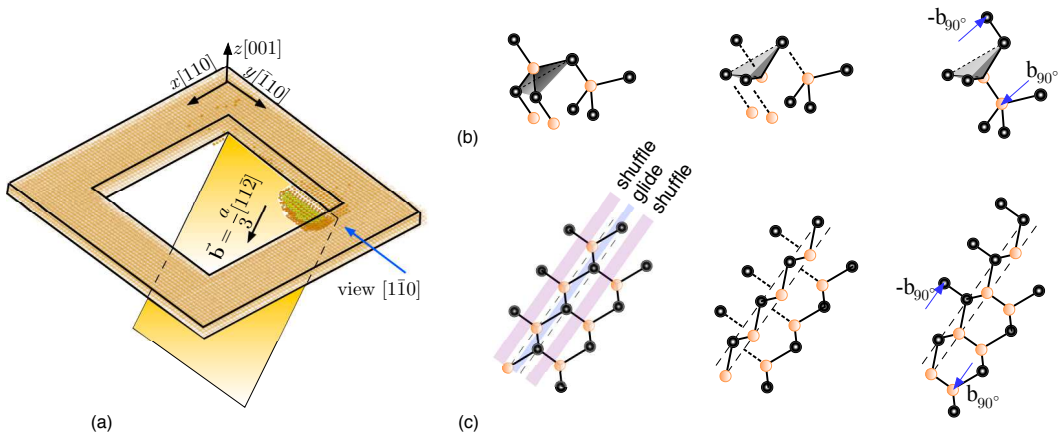}
\caption{Nucleation of shuffle-glide dislocation complex in MD. (a) Snapshot showing the nucleus from the corner of a sharply bent surface step. 
Only the atoms surrounding the surface step and those enclosed by the dislocation loop are shown.
(b) 3-d view and (c) 2-d view from $[1\bar{1}0]$ direction of the shuffle-glide complex slip mechanism.  
The shaded regions in (c) illustrates the shuffle-set planes between more widely separated atomic layers and the glide-set plane between more closely spaced atomic layers.
The Burgers vector of the dislocation, $\vec{\mathbf{b}}=\frac{a}{3}[{1}1\bar{2}]$, is split into two Shockley partial Burgers vectors, $\vec{\mathbf{b}}= \vec{\mathbf{b}}_{90^\circ} + \vec{\mathbf{b}}_{90^\circ}$, where $\vec{\mathbf{b}}_{90^\circ} =\frac{a}{6}[{1}1\bar{2}]$. The two Shockley partial dislocations are located on the shuffle-set planes immediate above and below the glide-set plane containing the stacking fault. }
\label{fig:snapshots}
\end{minipage}
\end{figure*}

Fig.~\ref{fig:neb-data}(c) shows the predicted energy barriers for shuffle-set ($60^\circ$ perfect) dislocation nucleation from both straight and sharply bent surface steps. 
Since such steps are only a few atomic layers high, it is assumed that surface features like these occur naturally during the growth of the semiconductor film.
The energy barrier calculations are performed using a modified string method which has demonstrated improved numerical stability even at high stresses (see {supplementary information} for more details).
The energy barrier calculations require an initial state, which is a dislocation-free Si crystal, and a final state, which contains a shuffle-set dislocation half loop on the $({1}11)$ plane with a Burgers vector of $a/2[10\bar{1}]$. 
The calculations are performed with external strain of increasing magnitudes to reduce the activation energy. 
The dashed line in Fig.~\ref{fig:neb-data}(c) corresponds to the energy barrier for nucleation from a straight surface step.
The extrapolation of this curve to the strain where the energy barrier vanishes gives the critical nucleation strain, which is about $7.8\%$ here.
This is consistent with the lowest critical strain value previously reported in the literature (Fig.~\ref{fig:literature_summary}).
%

%
%
The solid lines with open symbols in Fig.~\ref{fig:neb-data}(c) correspond to the nucleation barrier for the shuffle-set dislocation from a sharply bent surface step, predicted by the SW and MEAM potentials.
%
They are significantly lower than the nucleation barrier from the straight step, leading to a critical nucleation strain of $6.4\%$ (predicted by the SW potential) corresponding to point L in Fig.~\ref{fig:literature_summary}.
%
%
This result suggests that sharply bent surface steps may induce local stress concentrations that lower the nucleation barrier more effectively than straight steps.
%
This finding is significant in light of the large number of previous calculations of dislocation nucleation barrier in Si from various 2D surface features~\cite{godet2009evidence, izumi2008dislocation, shima2010reaction, thaulow2011atomistic, li2010dislocation, godet2004computer, saeed2009transition, godet2006dislocation} with counterintuitive findings.  For example, the barrier from re-entrant corners~\cite{izumi2008dislocation,shima2010reaction}, which are considered as strong stress concentrators, is even higher than that from a flat surface.

In order to confirm the significant effect of bent surface steps, we perform finite temperature MD simulations.
This is important because energy barrier calculations require the assumptions of what dislocation types to be nucleated (final state) as well as an initial guess of the nucleation pathway.
Incorrect assumptions can contribute to the discrepancy with experiments. 
On the other hand, MD simulations, despite their severe time scale limit, do not require such assumptions.
%
%

%
We perform MD simulations with LAMMPS~\cite{plimpton1995fast} using the Verlet integrator with a time step of $1$ ${\rm fs}$ under the NVT ensemble through the Nos{\'e}-Hoover thermostat. 
%
Compressive strain is applied along the $[{1}10]$ direction starting at 4.6\% with increments of $0.1\%$ after every 4~ns until dislocation nucleates. 
Dislocation nucleation from the corner of the bent surface step is consistently observed at $5.6\%$ and $5.3\%$ strain for the SW and MEAM potentials, respectively, which are even lower than the predictions from energy barrier calculations above.
%
%
%

Surprisingly, the dislocation nucleated from the MD simulations is not a conventional shuffle-set dislocation, as assumed in our energy barrier calculations. 
Fig.~\ref{fig:snapshots}(a) shows that the nucleated dislocation loop contains a stacking fault area, while detailed analysis reveals that slip occurs on the shuffle-set planes (Fig.~\ref{fig:snapshots}(c)) on which stacking fault cannot exist.
The Burgers circuit analysis shows that the dislocation Burgers vector is $\vec{\mathbf{b}} = \frac{a}{3}[{1}1\bar{2}]$ and is split into two Shockley partial Burgers vectors each having the Burgers vector $\vec{\mathbf{b}}_{90^\circ} = \frac{a}{6}[{1}1\bar{2}]$.
The two Shockley partial dislocations actually exist on the two shuffle-set planes adjacent to the glide-set plane of the stacking fault.
This structure contradicts the conventional notion that partial dislocations cannot exist on shuffle-set planes.

%
%
%
%

Fig.~\ref{fig:snapshots}(b) and (c) illustrate the slip mechanism corresponding to this dislocation. 
The shaded areas (forming a pyramid) in Fig.~\ref{fig:snapshots}(b) highlight the atomic arrangements adjacent to the glide-set plane; the two atomic layers switch their positions relative to the glide-set plane (inverting the pyramid), forming a stacking fault.
Such a transformation results in a larger separation between these atoms and their neighbors across the shuffle-set plane, effectively breaking the bonds (changing from solid to dashed lines from the left to middle panel in Fig.~\ref{fig:snapshots}(b) and (c)).
The bonds across the shuffle-set planes are eventually restored after the atomic layers slide by $\vec{\mathbf{b}}_{90^\circ}$ relative to each other across the two shuffle-set planes.
%
%
%
This slip nucleation mechanism was first reported by~\cite{godet2004computer} as a ``micro-twin'', which nucleates from a straight surface step at 7.8\% strain. 
Because this dislocation consists of two Shockley partial dislocations on the shuffle-set plane and a stacking fault on the glide-set plane, here we will refer to it as a ``shuffle-glide dislocation complex''.
%
%

%

To re-establish consistency between MD and energy barrier calculations, we calculate the energy barrier for the nucleation of shuffle-glide dislocation complex from the bent surface step.
%
%
The filled symbols in Fig.~\ref{fig:neb-data}(c) correspond to the predictions from SW and MEAM potentials, respectively.
The critical strains, $5.6\%$ (SW) and $5.3\%$ (MEAM), are now fully consistent with the predictions from MD simulations.
%
%
%
%
Because the shuffle-glide dislocation complex is predicted to be the preferred nucleation mechanism (i.e. with the lowest critical strain) by both SW and MEAM potentials, it is less likely to be an artifact of the SW potential as previously suspected~\cite{godet2004theoretical}, and may actually exist in Si and induce subsequent nucleation of more common types of dislocations.
%
The combination of bent surface steps and shuffle-glide dislocation complex leads to a prediction of a critical nucleation strain of 5.3\%, much closer to the experimental value ($\sim 3\%$) than previous predictions.

%
%
%
In summary, we show that the discrepancy between experiments and atomistic models of dislocation nucleation in Si can be significantly reduced by considering three dimensional surface features such as sharply bent surface steps.  The nucleation of shuffle-glide dislocation complex at the bent step has the lowest critical strain, and is confirmed by both SW and MEAM potentials. Surface reconstruction is an important aspect for obtaining consistent results leading to these findings. 
These results suggest that other surface features and previously unexplored dislocation types may hold the key for the full resolution of this long-standing discrepancy.

We thank Dr. Amin Aghaei for helping set up the MD simulations in Lammps. This work is supported by Samsung SSI (X.Z) and supported by the U.S. Department of Energy, Office of Basic Energy Sciences, Division of Materials Sciences and Engineering under Award No. DE-SC0010412 (W.C.).
%
%
%

\bibliography{blackout}

\end{document}